\documentclass [twocolumn,a4,10pt]{article}
\usepackage{graphicx}
\usepackage{amssymb}
\usepackage{amsmath}
\usepackage{cite}
\numberwithin{equation}{section}

\newcommand{\captionfonts}{\small}

\makeatletter  
\long\def\@makecaption#1#2{%
  \vskip\abovecaptionskip
  \sbox\@tempboxa{{\captionfonts #1: #2}}%
  \ifdim \wd\@tempboxa >\hsize
    {\captionfonts #1: #2\par}
  \else
    \hbox to\hsize{\hfil\box\@tempboxa\hfil}%
  \fi
  \vskip\belowcaptionskip}
\makeatother   

\begin{document}
\begin{center}
{\Large Surface Superconductivity in Niobium for Superconducting RF Cavities}\\
S. Casalbuoni$^{1,2}$, E.-A. Knabbe$^2$, J. K\"otzler$^1$, L. Lilje$^2$,\\
 L. von Sawilski$^1$, P. Schm\"user$^3$, B. Steffen$^{2,3}$\\
\end{center}
$^1$Institut f\"ur Angewandte Physik, Universit\"at Hamburg\\
$^2$Deutsches Elektronensynchrotron DESY, Hamburg\\
$^3$Institut f\"ur Experimentalphysik, Universit\"at Hamburg\\





\section*{Abstract}
A systematic study is presented on the superconductivity (sc)
parameters of the ultrapure niobium used for the fabrication of
the nine-cell 1.3 GHz cavities for the linear collider project
TESLA. Cylindrical Nb samples  have been subjected to the same
surface treatments that are applied to the TESLA cavities:
buffered chemical polishing (BCP), electrolytic polishing (EP),
low-temperature bakeout (LTB). The magnetization curves and the
complex magnetic susceptibility have been measured over a wide
range of temperatures and dc magnetic fields, and also for
different frequencies of the applied ac magnetic field. The bulk
superconductivity parameters such as the critical temperature
$T_c=9.26~$K and the upper critical field $B_{c2}(0)=410 ~$mT are
found to be in good agreement with previous data. Evidence for
surface superconductivity at fields above $B_{c2}$ is found in all
samples. The critical surface field exceeds the Ginzburg-Landau
field $B_{c3}=1.695 B_{c2}$ by about 10\% in BCP-treated samples
and increases even further if EP or LTB are applied. From the
field dependence of the susceptibility and a power-law analysis of
the complex ac conductivity and resistivity the existence of two
different phases of surface superconductivity can be established
which resemble the Meissner and Abrikosov phases in the bulk: 
(1)~``coherent surface superconductivity'', allowing sc shielding
currents flowing around the entire cylindrical sample, for
external fields $B$ in the range $B_{c2}<B<B_{c3}^{coh}$ , and 
(2)~``incoherent surface superconductivity'' with disconnected sc
domains
 for  $B_{c3}^{coh}<B<B_{c3}$. The
``coherent'' critical surface field separating the two phases is
found to be $B_{c3}^{coh}=0.81 \,B_{c3}$ for all samples.
 The exponents in the power law analysis are different
for BCP and EP samples, pointing to different surface topologies.

\section{Introduction}

The proposed linear electron-positron collider project TESLA is based
on superconductor technology for particle acceleration. For a
centre-of-mass energy of 500~GeV an accelerating field of 23.4~MV/m
is required in the 1.3 GHz nine-cell cavities which are made from pure
niobium and cooled by superfluid helium at 2~K.
The cavities for the TESLA Test Facility
(TTF) linac are fabricated from 2.8 mm thick niobium sheets by deep
drawing and electron beam welding. A damage layer of about 150~$\mu$m
thickness is removed from the inner surface to obtain optimum
performance in the superconducting state. For the TTF cavities this
has been done so far by chemical etching which
consists of two alternating processes: dissolution of the natural
Nb$_2$O$_5$ layer by HF and re-oxidation of the niobium by a strongly
oxidizing acid such as nitric acid (HNO$_3$) \cite{gmelin_nb, kneisel_kfk}. To reduce the
etching speed a buffer substance is added, for example phosphoric acid,
and the mixture is cooled below $15^\circ$C. The standard procedure
with a removal rate of about 1 $\mu$m per minute is called buffered
chemical polishing (BCP) using an acid mixture of HF (40\%), HNO$_3$ (65\%) and H$_3$PO$_4$ (85\%) in a volume ratio of 1:1:2.
In the most recent industrial
production of 24 TTF cavities an average gradient $26.1 \pm 2.3~$ MV/m
at a quality factor $Q_0=1 \cdot 10^{10}$ was achieved. The technology
developed for TTF is hence adequate for TESLA-500 but considerable
improvements are needed for an upgrade of the collider to 800 GeV. A
detailed description of the present status of the nine-cell cavity
layout, fabrication, preparation and tests can be found in
\cite{tesla_cavities}.

\noindent After many years of
intensive R\&D there exists now compelling evidence that the BCP
process limits the attainable field in multi-cell niobium cavities to
about 30~MV/m. This is significantly below the physical limit of about
45~MV/m which is given by the condition that the radio frequency (rf) magnetic field has
to stay below the critical field of the superconductor. For the type
II superconductor niobium the maximum tolerable rf field appears to be
close to the thermodynamic critical field ($B_c\approx 190$~mT at 2 Kelvin).

\noindent An alternative surface preparation method is electrolytic
polishing (EP). The material is removed in an acid mixture (for
example HF and H$_2$SO$_4$) under the
flow of an electric current. Sharp edges or tips are smoothed out and
a very glossy surface can be obtained. Using electrolytic polishing,
scientists at the KEK laboratory in Tsukuba (Japan) achieved gradients
of up to 40 MV/m in single-cell cavities \cite{saito_superiority,
 saito_99}. Meanwhile gradients of 35 - 40 MV/m have been obtained
repeatedly in many 1.3~GHz single-cell test cavities
\cite{kako_99,lilje_99, lilje_03}. Recently the EP technology has
been successfully transferred to the nine-cell TESLA cavities yielding
a record value of 39 MV/m in a multicell cavity \cite{lilje_03_2}.

\noindent The superiority of EP as compared to BCP can be
partially understood in terms of the much reduced surface roughness. 
The sharp ridges at the grain boundaries of an etched niobium surface may 
cause local enhancements of the rf magnetic field and
thereby lead to a premature breakdown of superconductivity at
these localized spots. A model based on this idea, developed by
Knobloch et al. \cite{knobloch}, is able to explain the reduction of the
quality factor at high field. However, a puzzling observation which
does not fit into this geometrical picture was made during the
CERN-DESY-Saclay R\&D programme on the electropolishing of single-cell
cavities~\cite{lilje_03}: after the EP and rinsing
with ultrapure water the cavities failed to reach full performance but
exhibited a strong decrease of
quality factor when high fields were approached. Applying a 24 - 48
hour ``bakeout'' at 120-140$^\circ$C to the evacuated cavity resulted
in a dramatic improvement: very high
gradients were accessible and the ``Q drop'' vanished. It should be
noted that the EP treatment at KEK \cite{saito_superiority,
 saito_99} already included a bakeout at $85^\circ$C.
The BCS surface resistance at 1.3-1.5~GHz is found to be
 reduced by a factor of 2 after baking~\cite{kneisel_99, lilje_03}.
However, by removing a layer of $\sim 200-300~\mbox{nm}$ in steps
of $\sim 50~\mbox{nm}$ from the EP surface by ``oxipolishing'' the
reduction in $R_{BCS}$ is lost and the ``Q drop''
reappears~\cite{kneisel_99}. The reduction of $R_{BCS}$ during
baking has been attributed to oxygen atoms diffusing either from
the dielectric Nb$_2$O$_5$ layer or from intergrain niobium
oxides/suboxides down to a depth of 200-300~nm~\cite{kneisel_99}.
 For further discussions we refer to the
review talks by P. Kneisel \cite{kneisel_03} and B. Visentin
\cite{visentin_03} at the SRF2003 Workshop. Relevant
investigations on the influence of surface treatment on
single cell cavities have been presented at the same workshop by Ciovati et al.~\cite{ciovati}.\\

\noindent Another explanation for the superiority of electropolished cavities
has been proposed by Halbritter \cite{halbritter99}.
The smoother surface and the cleaner
grain boundaries of an EP-treated cavity
 may lead to a reduction of dielectric surface losses
which are caused by interface tunnel exchange between the conduction electrons
in Nb and localized states in the Nb-Nb$_2$O$_{5-x}$
interface.\\

\noindent Magnetic measurements on niobium samples are a useful tool to explore
the surface treatments which improve cavity performance. This idea is
based on the fact that for pure niobium the ratio
$\kappa=\lambda_L/\xi$ is in the order of unity, so that surface
superconductivity and electromagnetic losses of microwave fields reside
in thin surface sheaths of nearly the same thickness, given by the
correlation length $\xi$ and the penetration depth $\lambda_L$,
respectively.
The experimental studies on the magnetization and
susceptibility of niobium samples presented here have been carried out
with the aim to gain an understanding of the
superconducting properties in this sheath and
their dependence on the various treatments (BCP, EP, bakeout) that
are applied to the TESLA cavities. The samples have been cut from
remainders of the niobium sheets used in cavity production and have
been subjected to the same
chemical, electro-chemical and thermal treatments as the TESLA cavities.

\section{Experimental Procedure}

The samples for the magnetization and susceptibility measurements 
are cylinders with a raw diameter of 2.5~mm and a height of 2.8~mm 
which are cut by electro-erosion from the Nb sheets used for rf cavity
production. The niobium is of high purity with
 a residual resistivity ratio $RRR$ of
300. The impurity contents is given in Table \ref{Nb_spec}. 
The electroerosion leaves an oxide surface layer of
several $\mu\mbox{m}$ thickness  which is removed
 by chemical etching of about 50 $\mu\mbox{m}$, applying the standard BCP
process. The samples are rinsed with ultrapure
water and annealed for 2 hours at 800 $^{\circ}$C in a vacuum furnace
 ($p <10^{-7}$~mbar) to remove dissolved hydrogen and relieve
mechanical stress in the material. After the annealing a second
50~$\mu$m BCP and water rinsing is applied. In this state the samples
have a similar surface structure as the BCP-treated cavities. 

\begin{table}[hbt]
\begin{center}
\caption{Impurity contents of the Nb samples.}
\vskip 0.2truecm
\begin{tabular}{|c|c||c|c|}
\hline Ta & 210-950 ppm & O & $<$ 40 ppm\\
\hline W & $<$ 100 ppm & N & $<$ 20 ppm\\
\hline Mo & $<$ 50 ppm & C & $<$ 20 ppm\\
\hline Ti & $<$ 40 ppm & H & $<$ 3 ppm\\
\hline Fe & $<$ 30 ppm &&\\
\hline
\end{tabular}
\label{Nb_spec}
\end{center}
\end{table}

\noindent Several samples have been electropolished after the BCP treatment. 
Only the cylindrical surface has been polished with a removed
thickness between $40~\mu\mbox{m}$
and $165~\mu\mbox{m}$. The EP has been carried out at room
 temperature 
in a mixture of HF
($56\%$) and H$_2$SO$_4$ ($90\%$). Low temperature
baking (LTB) has been applied both to BCP- and EP-treated samples,
with a bakeout temperature of $100\pm 1^{\circ}\mbox{C}$,
$120\pm 5^{\circ}\mbox{C}$, $123\pm 1^{\circ}\mbox{C}$ resp. $144\pm
1^{\circ}\mbox{C}$ and a baking time between 12 and 96 hours in a vacuum
furnace at $p <10^{-7}$~mbar. Three of the BCP-treated and baked 
 samples have been subsequently etched by
 $1~\mu$m, $5~\mu$m and $10~\mu$m in order to investigate whether possible baking
effects are lost by the removal of the surface layer. 

\noindent Scanning Electron Microscopy and Atomic
Force Microscopy \cite{gil} reveal a very low
surface roughness of about $ 1~\mbox{nm}$ on the Nb grains
(area of about $10\times 10~\mu\mbox{m}^2$). The steps at grain
boundaries range from 1 $\mu$m to a few $\mu$m on
BCP-treated samples~\cite{lilje} and are below 100 nm after the EP. 

\noindent The sample magnetization is determined with a commercial
SQUID magnetometer (Quantum Design MPMS$_2$) at temperatures
ranging from 2~K to 300~K in external dc fields between zero and
1 Tesla. The
SQUID magnetometer allows also to measure the ac susceptibility
at 10~Hz. A second magnetometer, based on the mutual inductance
technique, is used in combination with a 
lock-in amplifier (EG\&G PARC Model 5302) to extend the ac
susceptibility measurements up to frequencies of 1~MHz in the
temperature range between 1.5~K and 4.2~K (pumped $^4$He cryostat).
The linearity of the response to the applied ac field 
has been verified for ac field amplitudes from $0.1 ~\mu$T to $500~\mu$T. 
In all measurements the
external magnetic fields (dc and ac) are carefully aligned
parallel to the symmetry axis of the cylinders, so that above
$H_{c2}$ surface superconductivity can only nucleate on the
cylindrical walls and not on the end faces. The demagnetization factor
$N_Z=0.36$ derived from the initial slope of the dc magnetization curve
$M(H)=-H/(1-N_Z)$ agrees with the theoretical expression
$N_Z=1-1/(1+qa/b)$~\cite{Brandt2}. Here
$q=4/3\pi+2/3\pi$~tanh$(1.27b/a$~ln$(1+a/b))$, where $h=2b$ is the
height and $a$ the radius of the cylinder. The magnetization and
susceptibility data presented in the next sections have been corrected
using this demagnetization factor.
\section{Bulk superconductivity}

\begin{figure}[htb]
\centering
\includegraphics[width=80mm]{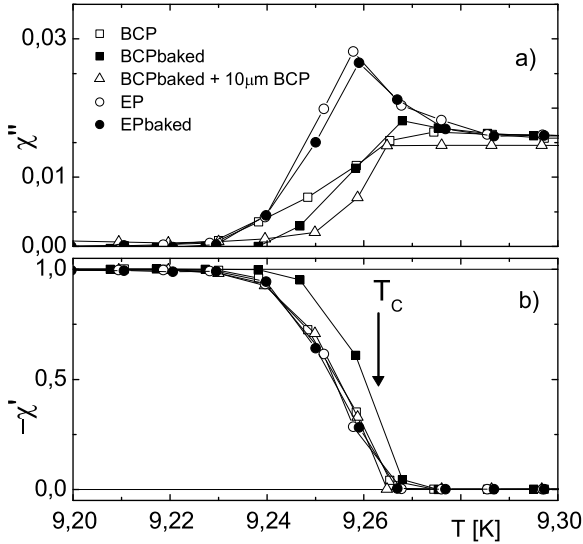}
\caption{(a) Imaginary and (b) real part of the linear
 ac susceptibility recorded near the zero-field transition
temperature of the Nb-cylinders under investigation. Note the
different $\chi''$-scale in (a).} \label{Tc}
\end{figure}

The superconducting transition temperature
$T_c$ is determined from the onset of the screening component $\chi'$
of the complex ac susceptibility $\chi=\chi'-i\chi''$, measured at vanishing
dc field as a function of
temperature (see Fig.~\ref{Tc}). The average critical temperature
$T_c=9.263\pm 0.003~\mbox{K}$ of all samples agrees with the value
$T_c=9.25\pm 0.01~\mbox{K}$ reported by Finnemore at
al.~\cite{finne} for high purity Nb ($RRR=1600\pm 400$).

\begin{figure}[hbt]
\centering
\includegraphics[width=80mm]{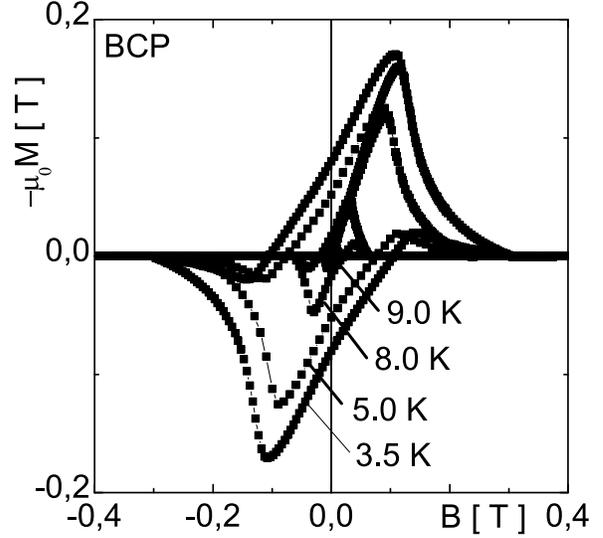}
\caption {Magnetization isothermal loops for a BCP sample. Plotted
is $-\mu_{0}M$ as function of the applied field $B=\mu_{0}H$.}
\label{Magloop}
\end{figure}

\begin{figure}[htb!]
\centering
\includegraphics[width=80mm]{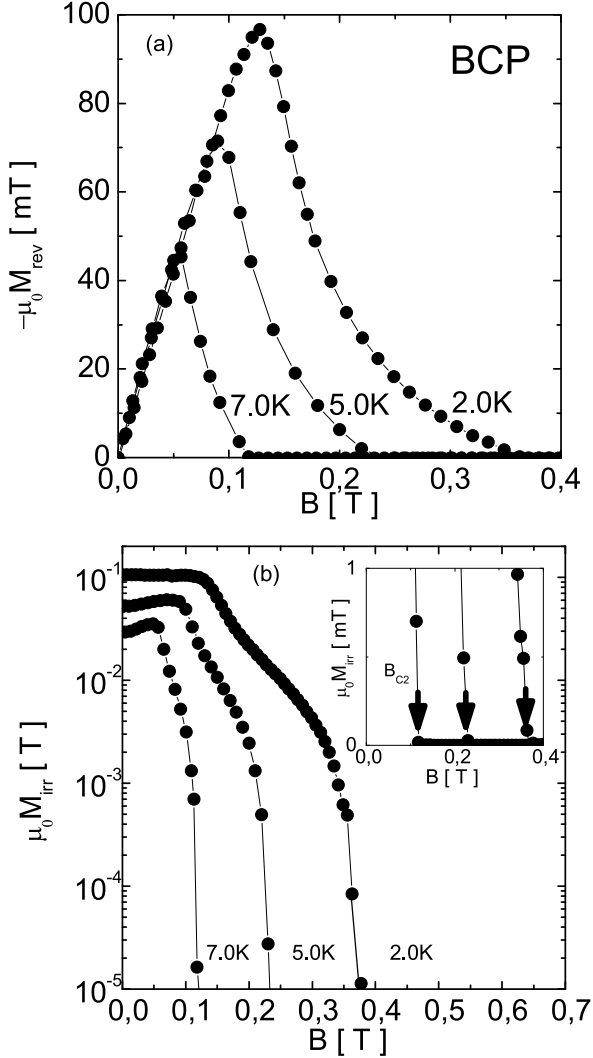}
\caption{(a) Reversible magnetization of a BCP sample at 2 - 7
K, evaluated from the isothermal loops. (b) Irreversible magnetization at 2, 5 and 7 K,
 showing the sharp onset of bulk superconductivity at $B_{c2}(T)
=\mu_0 H_{c2}(T)$.} \label{Mrev}
\end{figure}

\noindent The complex ac susceptibility\footnote{The sign of the imaginary
part refers to a time dependence $\exp(+i\omega t)$ of the
electromagnetic fields.}
$\chi(\omega)=\chi'(\omega)-i\chi''(\omega)$ is related to the
conductivity of a cylinder of radius $a$ by~\cite{maxwell, clem}
\begin{eqnarray}\label{chi-sigma}
\chi'(\omega)-i\chi''(\omega)&=&\frac{2I_1(x)}{xI_0(x)}-1 \\ \quad
{\rm with} \quad x&=&\sqrt{i\omega a^2\mu_0
(\sigma'(\omega)-i\sigma''(\omega))}\;. \nonumber
\end{eqnarray}
Here $I_0(x)$ and $I_1(x)$ are modified Bessel functions. For a
short cylinder with heigth $h=2a$ the radius must be replaced by
$a/\sqrt{2}$~\footnote{~The effect of sample geometry on the
relation between $\chi(\omega)$ and $\sigma(\omega)$ is
extensively studied in Ref. \cite{Brandt}.}. The complex
conductivity $\sigma(\omega)$ is obtained from the measured
susceptibility $\chi(\omega)$ by means of an inversion
routine~\cite{k}. The average Ohmic resistivity at 10 Hz and $T >
T_c$ is found to be $\rho_n=(0.5\pm 0.1)~$n$\Omega$m, confirming
the purity of the samples and the specification of the
manufacturers of $RRR\simeq 300$.

\begin{figure}[htb]
\centering \includegraphics[width=80mm]{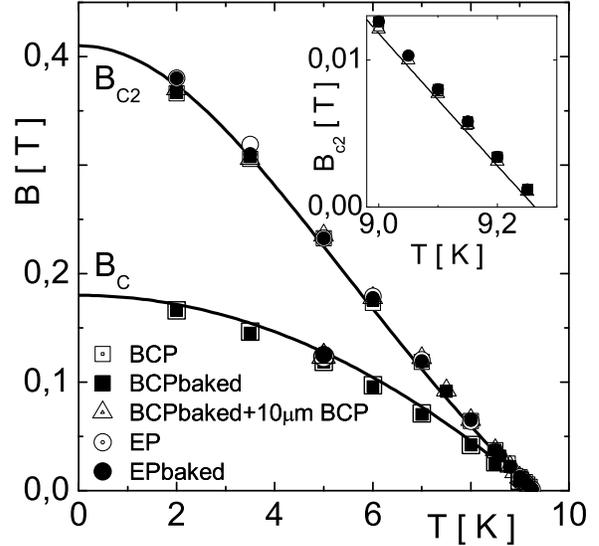}
\caption{Temperature variation of the upper critical field
$B_{c2}=\mu_0H_{c2}$ and of the thermodynamic critical field $B_c$ of
all cylinders. Solid curves: fits based on Eqs.~(\ref{eqHc}) and
(\ref{eqHc2}). The expected linear behaviour of $B_{c2}$ close to
$T_c$ is displayed in the inset.} \label{Hc2}
\end{figure}

\begin{table}[hbt]
\begin{center}
\caption{Parameters of the Nb samples.}
\vskip 0.2truecm
\begin{tabular}{|l|c|c|}
\hline 
&BCP&EP \\ \hline 
$T_c$~[K] &\multicolumn{2}{c|}{$9.263\pm
0.003$} \\ \hline 
$RRR$& \multicolumn{2}{c|} {$\approx 300$} \\ \hline 
surf. roughness& \multicolumn{2}{c|} {$$}\\ 
on grain [nm] &\multicolumn{2}{c|} {$\approx 1$} \\ \hline 
steps at grain bound.&
1-5~$\mu$m & $\lesssim 0.1\mu$m \\ \hline 
$B_c(0)~$[mT]&\multicolumn{2}{c|} {$180\pm 5$} \\ \hline 
$B_{c2}(0)~$[mT] &\multicolumn{2}{c|} {$410\pm 5$} \\ \hline
$J_c(0,0)~[\mbox{A/mm}^2]$&$240\pm 10$ &$180\pm 10$ \\ \hline
\end{tabular}
\label{bulk}
\end{center}
\end{table}

\noindent In order to determine the thermodynamic critical field
$H_c$ the reversible magnetization $M_{rev}=(M_{+}+M_{-})/2$ is
deduced from the hysteresis loops (Fig.~\ref{Magloop}), where
$M_+$ corresponds to the ascending and $M_-$ to the descending
branch of the loop. The reversible magnetisation of a BCP sample 
is plotted in Fig.~\ref{Mrev}a. The thermodynamic
critical field is given by
\begin{equation}
 H^2_{c}=2\mu_0 \int_0^{H_{c2}}M_{rev}(H) dH.
\end{equation}
 The critical field at $T=0$ is derived using the empirical law
\begin{equation}
H_{c}(T)=H_{c}(0)(1-(T/T_c)^2) \label{eqHc}
\end{equation}
where $T_c$ is given in Tab.~\ref{bulk}. The upper critical field
$H_{c2}$ can be determined with high accuracy
from the onset of the irreversible
magnetization $M_{irr}=(M_{+}-M_{-})/2$, as shown in the inset of
Fig.~\ref{Mrev}b.
The $H_{c2}$ data are well described by the temperature
dependence~\cite{tinkham1}
\begin{equation}
H_{c2}(T)=H_{c2}(0)\frac{(1-(T/T_c)^2)}{(1+(T/T_c)^2)}\;.
\label{eqHc2}
\end{equation}
The thermodynamic and upper critical fields\footnote{In this
paper we use the SI unit system and quote magnetic fields in the
form $B=\mu_0 H$ with the unit Tesla.} are plotted in
Fig.~\ref{Hc2} as functions of temperature.
 Averaged over all samples we get
$B_{c2}(0)=\mu_0 H_{c2}(0)=410\pm 5~$mT and $B_c(0)=\mu_0 H_{c}(0)=180\pm
5~$mT (see also Tab.~\ref{bulk}). While
$B_{c2}(0)$ agrees with the upper critical field reported in Ref.~\cite{finne} for
pure niobium, our $B_c(0)$ is 10\% smaller than the value $B_c(0)=199\pm 1~$mT
quoted in~\cite{finne}.

\noindent Within the Ginzburg-Landau (GL) theory it
is possible to determine the GL parameter at zero temperature
$\kappa(0)=B_{c2}(0)/(\sqrt{2} B_{c}(0))=1.61\pm 0.07$,
the GL coherence length
$\xi(0)=\sqrt{\hbar/(2 e B_{c2}(0))}=28.3\pm 0.2~\mbox{nm}$
and the London
penetration depth $\lambda_L(0)=\kappa(0) \xi(0)=46\pm 2~\mbox{nm}$.
The data from differently treated samples are in good agreement and
confirm the expectation that
neither electropolishing nor low-temperature bakeout change the
superconductor parameters of the bulk niobium.

\noindent The hysteresis loops observed in the magnetization measurements are a
 clear proof of magnetic flux pinning. Under the assumption that
 pinning in the bulk dominates one can derive the critical current
 density from the irreversible part of the magnetization:
 $J_c=3 M_{irr}/a$, where $a$ is the radius of the Nb cylinder.
 We obtain $J_c=240 $ A/mm$^2$ for the BCP samples and
$J_c=180 $ A/mm$^2$ for the EP
 samples (both at $T=4.2~$K and $B=0$). Since EP does not affect the
 bulk this difference implies that surface flux pinning
 plays an important role too. Similar observations were made by De Sorbo
\cite{desorbo} who found large variations of $J_c$ between 10 and 1000
 A/mm$^2$ for Nb with $RRR \simeq 500$, depending on the mechanical
 and thermal treatment.

\section {Surface superconductivity}

\subsection{Critical surface field}

\begin{figure}[hbt]
\centering
\includegraphics*[width=71mm]{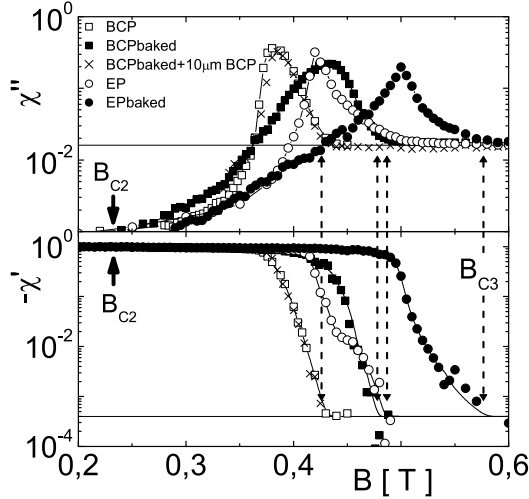}
\caption{Field dependence of the ac susceptibility on a
logarithmic scale of Nb cylinders with different surface
conditions. The data have been taken at
$T=5.0~$K, $f=10~\mbox{Hz}$ and $B_{ac}= 1~\mu$T. The
surface nucleation fields $B_{c3}$ (arrows) are defined by
the onset of an excess absorption ($\chi''-\chi''_n$) or excess shielding
($\chi'-\chi'_n$), where $\chi''_n$ resp. $\chi'_n$ are the imaginary
resp. real part of $\chi$ in the normal state (indicated by the
horizontal solid lines). } \label{Hc3all}
\end{figure}

\begin{figure}[hbt]
\centering
\includegraphics*[width=71mm]{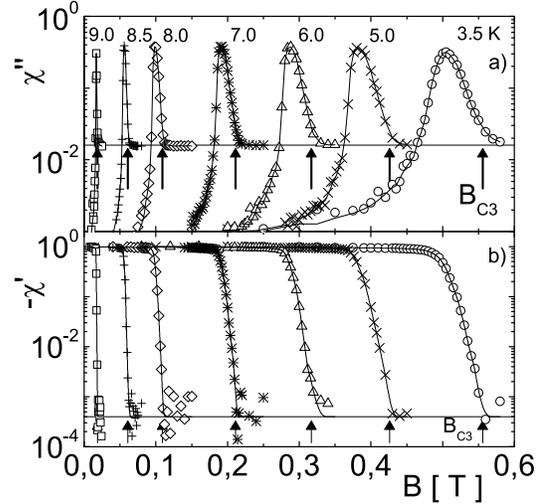}
\caption{(a) Imaginary and (b) real part of the ac
susceptibility of a BCP-treated sample as a function of field for temperatures
between 3.5~K and 9.0~K.}
 \label{Hc3BCP}
\end{figure}

\begin{figure}[hb!]
\centering
\includegraphics*[width=71mm]{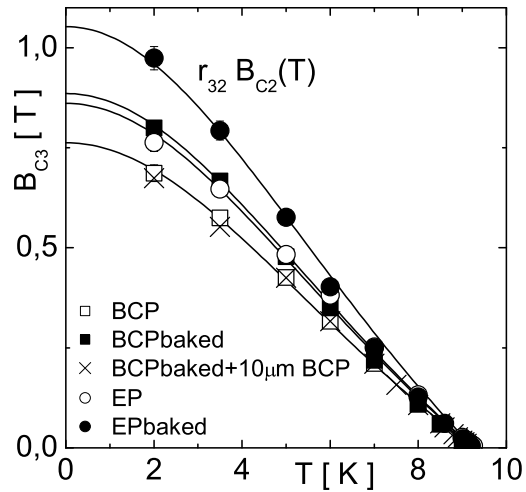}
\caption{Temperature dependence of critical surface field
 $B_{c3}=\mu_0 H_{c3}$ of the samples shown in
 Fig.~\ref{Hc3all}. The data are fitted to the
function $B_{c3}(T)=r_{32} B_ {c2}(T)$, where the ratio
 $r_{32}$ depends on the
surface preparation but not on temperature. This means that $B_{c2}$
 and $B_{c3}$ have the same temperature dependence.} \label{Hc3T}
\end{figure}

Close to the surface a type II superconductor remains in the
superconducting state even if the upper critical field of the bulk
is exceeded. According to the Ginzburg-Landau theory the
critical surface field is given by $B_{c3}=1.695
B_{c2}$~\cite{Stjames}.
 Owing to the high sensitivity of the SQUID magnetometer, evidence for surface
superconductivity in the field range $B_{c2}<B<B_{c3}$ is in
fact observable in all samples. In
Fig.~\ref{Hc3all} we show the complex ac susceptibility
$\chi'-i\chi''$ at 5.0~K and 10~Hz as a function
of the applied dc field for samples with different surface treatments:
BCP, EP, baked and unbaked.

\noindent Approaching the transition to superconductivity from the
normal state, i.e.~from high fields, the nucleation field $B_{c3}$
is commonly defined \cite{hopkinsfinnemore, rollinssilcox} by the
onset of screening, i.e., the appearance of a finite $|\chi'|$
above the suceptibility value $\chi'_n$ in the normal state. We 
observe a strong dependence of $B_{c3}$ on the surface condition: 
the BCP sample has the lowest $B_{c3}$, and both EP and baking increase 
the critical surface field.

\begin{figure}[htb]
\centering
\includegraphics*[width=80mm]{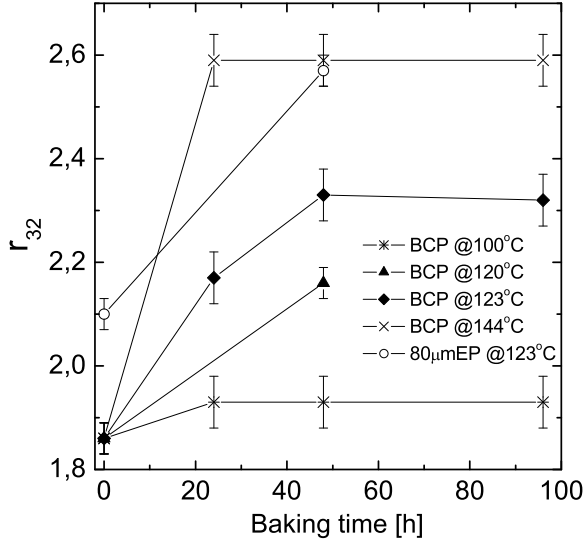}
\caption{The ratio $r_{32}=B_{c3}/B_{c2}$ as a function of baking
time
 for BCP and EP samples at different baking temperatures.}
\label{rbakingEP}
\end{figure}

\noindent The ac susceptibility
of a BCP sample at temperatures between 3.5~K and 9.0~K is plotted in
 Fig.~\ref{Hc3BCP}.
Similar measurements have been performed for all samples.
The temperature dependence of $B_{c3}$ for samples with different
surface treatment is shown in
Fig.~\ref{Hc3T}. The data are well described by the relation
\begin{equation}
B_{c3}(T)=r_{32}\cdot B_{c2}( T)
\end{equation}
 where the ratio $r_{32}$ depends on the surface
preparation but not on the temperature, see Tab.~\ref{r} and
Fig.~\ref{rbakingEP}.

\begin{table}[hbt]
\begin{center}
\caption{The ratio $r_{32}=B_{c3}/B_{c2}$ for all samples.} \vskip
0.2truecm
\begin{tabular}{|l|c|}
\hline Sample & $r_{32}=B_{c3}/B_{c2}$ \\
 \hline BCP only &$1.86\pm 0.03$ \\
 \hline\multicolumn{2}{|c|}{BCP+Baking}\\
 \hline BCP+24h 100$^\circ$C &$1.93\pm 0.05$ \\
 \hline BCP+48h 100$^\circ$C &$1.93\pm 0.05$ \\
 \hline BCP+96h 100$^\circ$C &$1.93\pm 0.05$ \\
 \hline BCP+48h 120$^\circ$C &$2.16\pm 0.03$ \\
 \hline BCP+24h 123$^\circ$C &$2.17\pm 0.05$ \\
 \hline BCP+48h 123$^\circ$C &$2.33\pm 0.05$ \\
 \hline BCP+96h 123$^\circ$C &$2.32\pm 0.05$ \\
 \hline BCP+24h 144$^\circ$C &$2.59\pm 0.05$ \\
 \hline BCP+48h 144$^\circ$C &$2.59\pm 0.05$ \\
 \hline BCP+96h 144$^\circ$C &$2.59\pm 0.05$ \\
 \hline\multicolumn{2}{|c|}{BCP+Baking+BCP}\\
 \hline BCP+48h 120$^\circ$C +1 $\mu m$ BCP &$1.90\pm 0.03$ \\
 \hline BCP+48h 120$^\circ$C +5 $\mu m$ BCP &$1.87\pm 0.03$ \\
 \hline BCP+48h 120$^\circ$C +10 $\mu m$ BCP &$1.86\pm 0.03$ \\
 \hline\multicolumn{2}{|c|}{BCP+EP}\\
 \hline BCP+40$\mu m$EP &$1.92\pm 0.05$ \\
 \hline BCP+80$\mu m$EP &$2.10\pm 0.03$ \\
 \hline BCP+145$\mu m$EP &$1.99\pm 0.05$ \\
 \hline BCP+165$\mu m$EP &$1.99\pm 0.05$ \\
 \hline\multicolumn{2}{|c|}{BCP+EP+Baking}\\
 \hline BCP+40$\mu m$EP+48h 123$^\circ$C &$2.64\pm 0.05$ \\
 \hline BCP+80$\mu m$EP+48h 123$^\circ$C &$2.57\pm 0.05$ \\
 \hline BCP+145$\mu m$EP+48h 123$^\circ$C &$2.40\pm 0.05$ \\
 \hline BCP+165$\mu m$EP+48h 123$^\circ$C &$2.40\pm 0.05$ \\
 \hline
\end{tabular}
\label{r}
\end{center}
\end{table}

\noindent The reproducibility of the measurements of the ratio $r_{32}$ has been
demonstrated on six BCP samples, two for each of the Nb producers
(all Nb sheets
have $RRR \approx 300$). The average value for the BCP samples is
$r_{32}=1.86 \pm 0.03$ and hence larger than
Ginzburg-Landau value $r_{GL}=1.695$.

\noindent If a BCP sample receives an electropolishing of $80~\mu \mbox{m}$
 the ratio
 $r_{32}$ grows by $\sim 12\%$.
Also a low temperature baking (LTB) increases the ratio
$r_{32}$ considerably, see Fig. \ref{rbakingEP}.
 Here the reproducibility has been demonstrated on
 six BCP samples, two for each of the Nb producers, which were baked for 48~h at
 $120^\circ \mbox{C}$ and yielded $r_{32}=2.33 \pm 0.05$.
In the bakeout at $123^\circ \mbox{C}$ the ratio
 $r_{32}$ grows with increasing baking time. At $144^\circ \mbox{C}$
a saturation value of
$r_{32}=2.59\pm 0.05$ is reached after 24~hours without further increase towards longer
 baking times. The gain in the ratio $r_{32}$ is completely lost if a
$1~\mu \mbox{m}$ surface layer is removed by chemical etching. This
proves that only a thin surface layer is modified by the bakeout.

\noindent Baking at $100^\circ \mbox{C}$ yields only a slight rise in $r_{32}$.
 Moreover, no appreciable change is observed by extending the
baking time from 24~h to 94~h. This means that $100^\circ$C is
 probably too low a temperature for changing the superconductor
 properties at the Nb surface significantly.\\

\noindent The existence of a critical surface field exceeding the
upper critical field of the bulk, $B_{c3}=r_{GL}\, B_{c2}$ with
$r_{GL}=1.695$, was predicted by de Gennes and Saint-James
 by solving the linearized Ginzburg-Landau equation
in the presence of a plane boundary which creates a mirror image
of the potential. If we retain this idea the increased ratio
$r_{32}>r_{GL}$ may be explained by assuming a larger value of
$B_{c2}$ close to the surface, $B_{c2}^{surf}
>B_{c2}^{bulk}$, and by relating $B_{c3}$ to this field:
 $B_{c3}=r_{GL}\,B_{c2}^{surf}$. The enhanced $B_{c2}^{surf}$ may be caused for
instance by a higher concentration of impurities
and a reduced coherence length.

\noindent One obvious contaminant is
oxygen. After the discovery of the bakeout effect in cavities
systematic studies have been conducted on the morphological and chemical
structure of niobium samples, for instance by surface-sensitive
methods such as XPS (X-ray induced photoelectron spectroscopy)
\cite{dacca_98, claire_99, polen_bake}. These studies indicate
that baking causes a partial disintegration of the external
Nb$_2$O$_5$ layer and the formation of metallic suboxides (NbO,
NbO$_2$).

\begin{figure}[htb]
\centering
\includegraphics*[width=65mm]{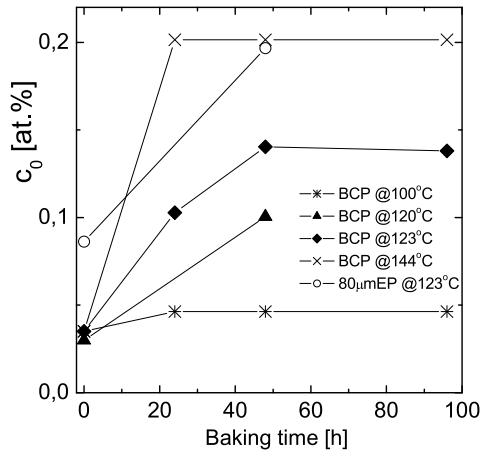}
\caption{Oxygen concentration as a function of baking time for different surface treatments.}
\label{co}
\end{figure}

\noindent The effect of interstitial oxygen on the upper critical field of
niobium was studied by Koch et al. \cite{koch_74}.
 The concentration $c_{\rm O}$ of oxygen in atomic
per cent can be derived from an expression valid at 4.2~K \cite{DasGupta}
$$c_{\rm O}~[at. ~\%]=~1.475 \cdot
 10^{-3}(B^{surf}_{c2}~ {\rm [mT]}-276)\;.$$
Under the assumption that the enhanced surface fields
$B_{c2}^{surf}$ are indeed caused by oxygen this formula is used to compute the oxygen concentration listed in Table~\ref{model}.
Figure~\ref{co} shows that the oxygen concentration increases if a BCP sample is subjected to EP or LTB.

\begin{table}[hbt]
\begin{center}
\caption{Parameters of the oxygen-contamination model and the Shmidt
 model.}
\begin{tabular}{|l|c|c|c|}
 \hline Sample & $c_{\rm O}$~(at.$\%$O)& $d_{min}~$[nm]&
 $\ell_{max}~$[nm]\\ \hline BCP only& 0.035& 2.5&436 \\
 \hline\multicolumn{4}{|c|}{BCP+Baking}\\ \hline 24h 100$^\circ$C &
 0.052& 3.5&318 \\ \hline 48h 100$^\circ$C & 0.052& 3.5&318 \\
 \hline 96h 100$^\circ$C & 0.052& 3.5&318 \\ \hline 48h
 120$^\circ$C & 0.106& 6.5&169 \\ \hline 24h 123$^\circ$C &
 0.109& 6.5&166 \\ \hline 48h 123$^\circ$C & 0.147& 8.5&125 \\
 \hline 96h 123$^\circ$C & 0.144& 8.5&127 \\
 \hline 24h 144$^\circ$C & 0.209& 12&90 \\
 \hline 48h 144$^\circ$C & 0.209& 12&90 \\
 \hline 96h 144$^\circ$C & 0.209& 12&90 \\
 \hline\multicolumn{4}{|c|}{BCP+Baking (48h at 120$^\circ$C)+ short BCP}\\
 \hline 1 $\mu$m BCP & 0.045& 3.&360 \\
 \hline 5 $\mu$m~ BCP & 0.037& 2.5&414 \\
 \hline 10 $\mu$m~ BCP & 0.035& 2.5&436 \\
 \hline\multicolumn{4}{|c|}{BCP+EP}\\
 \hline 40 $\mu$m~EP & 0.049& 3&331 \\
 \hline 80 $\mu$m~EP & 0.092& 6&192 \\
 \hline 145 $\mu$m~EP & 0.066& 4&259 \\
 \hline 165 $\mu$m~EP & 0.066& 4&259 \\
 \hline\multicolumn{4}{|c|}{BCP+EP+Baking (48h at 123$^\circ$C)}\\
 \hline 40 $\mu$m~EP & 0.221& 13&85 \\
 \hline 80 $\mu$m~EP & 0.204& 12&92 \\
 \hline 145 $\mu$m~EP & 0.163& 9.5&113 \\
 \hline 165 $\mu$m~EP & 0.163& 9.5&113 \\
 \hline
\end{tabular}
\label{model}
\end{center}
\end{table}

\noindent In a study on BCP cavities recently performed at Jefferson Lab,
Ciovati et al.~\cite{ciovati} measured an average surface
$RRR=206\pm 3$. Under the same assumption that oxygen is the main
contaminant they find $c_O=0.017$at.$\%$ which, considering the
different BCP conditions in the two laboratories, is in reasonable
agreement with our results for BCP samples of $c_O=0.035$at.$\%$.

\noindent Shmidt \cite{shmidt} has proposed a model in which
 the impurities are contained
in a layer whose thickness is less than the coherence length of the
bulk $d\leq\xi$~. This model predicts the following
relation:
\begin{equation}\label{Gorkov}
r_{32}=1.67\left(1+\left(1-\chi_G\right)\sqrt{1.7}\frac{d}{\xi(T)}\right).
\end{equation}
 Here $\chi_G$ is the Gor'kov impurity function which relates the GL $\kappa$ parameters
 in the pure bulk and
 the point defected surface, $\chi_G=\kappa/\kappa_{surf}$. Since
$(1-\chi_G )\leq 1$ and $d/\xi\leq 1$ the maximum ratio is
$r_{max} \leq 3.8$, which is consistent with our results. Eq.
(\ref{Gorkov}) embodies two unknowns, the mean free path $\ell$
and and the thickness $d$ along with
 $\xi_0=1.35\,\xi(0)$~\cite{gorkov}, hence we can
only consider limiting cases. In the dirty limit $\ell \rightarrow 0$
 ones gets $\chi_G \simeq 1.33\, \ell/\xi_0\rightarrow 0$ and
$r_{32}$ increases with increasing thickness $d$ of the contaminated
 layer.
 The minimum values of $d$ in the dirty limit are listed
in Tab.~\ref{model}. In the clean limit $\ell \gg \xi_0$ one has
$\chi_G \simeq 1-0.884 \,\xi_0/\ell$, hence the maximum value of
$\ell$ is given by putting $d=\xi$. The results are listed in
Tab.~\ref{model}. Both the oxygen contamination model and the
Shmidt model are consistent with the idea of oxygen diffusion
 during baking from the $\sim 5$~nm thin Nb$_2$O$_5$
sheath into deeper layers~\cite{kneisel_99}.

\section{Coherent surface superconductivity}

\subsection{Surface conductivity  and resistivity}
\begin{figure}[t]
\centering \includegraphics*[width=65mm]{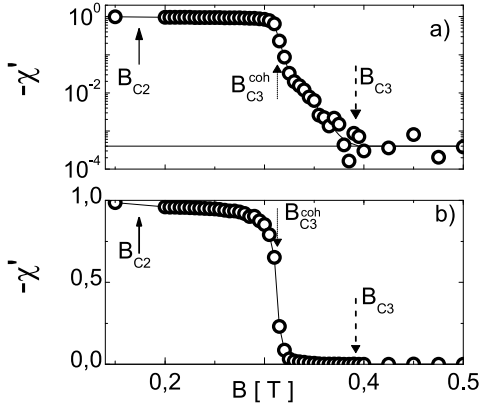}
\caption{a) The negative real part of the susceptibility of an EP cylinder at 6.0~K
and 10~Hz, plotted on a logarithmic scale, as a function of the
applied field. The start of the growth of $|\chi'|$ defines the
critical surface field $B_{c3}=\mu_0 H_{c3}$. b) The same data on a
linear scale. The field $B_{c3}^{coh}$ marks the onset of coherent
surface superconductivity with a strong growth of $|\chi'|$ towards
$|\chi'|=1$.} \label{chi_linlog}
\end{figure}

We now examine in more detail the behaviour of the niobium samples
for applied magnetic dc fields in the range between $B_{c2}$ and
$B_{c3}$  where only a thin surface sheath is superconducting
while the bulk is in the normal state. Coming down from high
fields, the screening part of $\chi$ starts to grow upon crossing
$B_{c3}$.
 Owing to the high sensitivity of the SQUID magnetometer, the onset of screening
 is clearly visible when $-\chi'$ is plotted on a
 logarithmic scale (Fig.~\ref{chi_linlog}). This allows a precise
 determination of the critical surface field $B_{c3}$. On a linear scale,
 however, the onset of surface superconductivity 
 is barely visible, while a strong rise of $-\chi'$ is observed at a lower field
 $B_{c3}^{coh}$. Following Ref.\cite{lars2}, we call this field the {\it coherent critical
 surface field}, for reasons explained below. It is still well above
 the nucleation field $B_{c2}$ of the bulk.\\

\begin{figure}[phbt]
\centering
\includegraphics*[width=65mm]{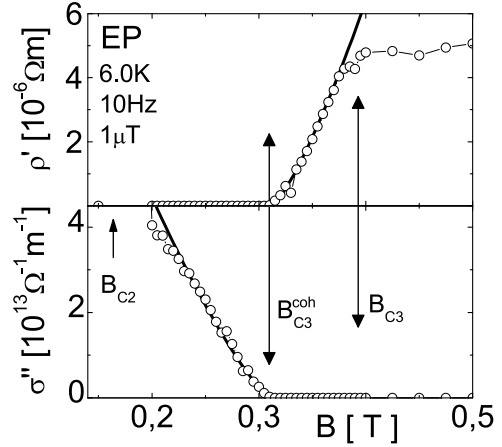}
\caption{Top: The real part of the resistivity
 $|\rho'|$ of an EP sample as a function of field. For $B>B_{c3}^{coh}$ the data are well described by the power law curve $\rho' \propto |B-B_{c3}^{coh}|^{1.3}$. Bottom: The imaginary part of the conductivity in comparison with the power law $\sigma'' \propto |B_{c3}^{coh}-B|^{1.3}$. } \label{EPpower1}
\end{figure}

\noindent To gain a deeper insight into the physical mechanisms in this
region it is instructive to study the field dependence of the
complex conductivity $\sigma=\sigma'-i \sigma''$ and the
resistivity $\rho=1/\sigma= \rho' +i \rho''$. These quantities can
be calculated from the ac susceptibility with the help of
Eq.~(\ref{chi-sigma}). In the range $B_{c3}>B>B_{c3}^{coh}$ the
resistivity $\rho'$ is sharply dropping with decreasing $B$, but
there is only very little screening ($\sigma''$ is extremely
small), as can be seen from Fig.~\ref{EPpower1}. This can be
interpreted in the following way: for applied fields
$B_{c3}>B>B_{c3}^{coh}$ the surface sheath is not uniformly
superconducting but only in disconnected domains such that
super-currents flowing around the entire Nb cylinder are impeded.
The situation changes dramatically when the applied field falls
below $B_{c3}^{coh}$: here the Ohmic resistivity $\rho'$ vanishes and the imaginary part
$\sigma''$ of the conductivity rises steeply with decreasing
field. Since $\omega \sigma''$ is proportional to the Cooper pair
density\footnote{In the two-fluid model of superconductivity, the
imaginary part of the ac conductivity is given by
$\sigma''=2n_ce^2/(m_e \omega)$ where $n_c$ is the Cooper pair
density, see for example \cite{PS}.}, it was proposed in
Ref.~\cite{lars2} to associate $B_{c3}^{coh}$ with the onset of
long-range superconductivity in the surface sheath. This
observation justifies the name coherent surface field. Slightly
below $B_{c3}^{coh}$ the susceptibility reaches the value
$\chi'=-1$ which is characteristic of complete shielding.

\noindent The singular behaviour of $\rho'$ and $\sigma''$ near
the transition to coherent surface superconductivity can be
described by power laws in $\left|B-B^{coh}_{c3}\right|$, see
Fig.~\ref{EPpower1}. Above the transition one gets
\begin{equation}
\rho'(B) \propto (B-B_{c3}^{coh})^s \quad {\rm for}\quad
B_{c3}^{coh} <B< B_{c3}
\end{equation}
 while below the transition
\begin{equation}
\sigma''(B)\propto (B_{c3}^{coh}-B)^t \quad {\rm for}\quad B_{c2}
<B <B_{c3}^{coh} \;.
\end{equation}

\begin{figure}[pthb]
\centering
\includegraphics*{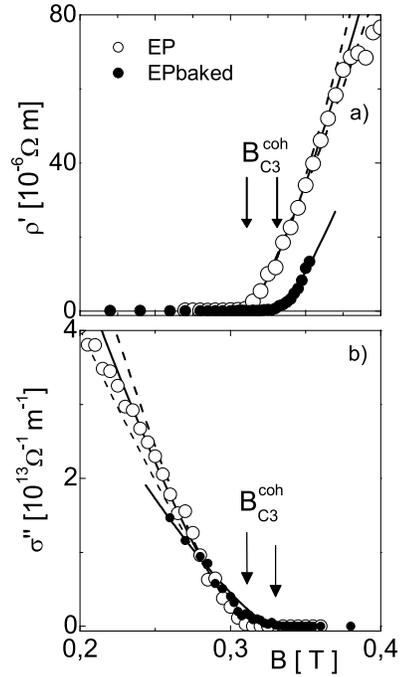}
 \caption{Analysis of the singular behaviour (a) of the real part of
  the resistivity $\rho'$ and (b) of the imaginary part of the conductivity
 $\sigma''$ of an electropolished Nb cylinder at 6 K. Open circles: 
before baking, closed circles: after baking. 
The solid curves are power-law fits revealing the same exponents before and after
baking. Note that the coherent surface field is larger after baking.}
 \label{EPpower2}
\end{figure}

\begin{figure}[pbht]
\centering
\includegraphics*{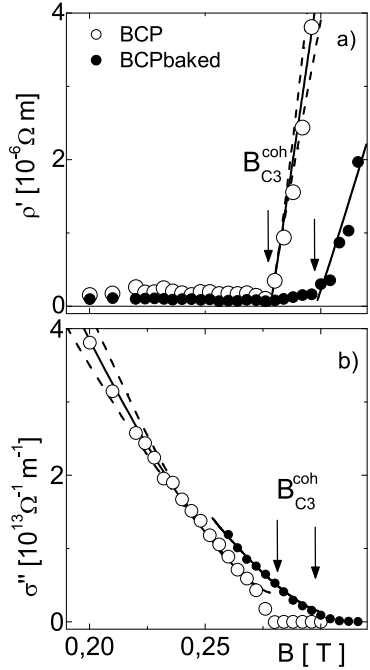}
 \caption{(a) The real part of the resistivity $\rho'$ and (b) the imaginary part of the conductivity
 $\sigma''$ of a chemically etched (BCP) cylinder. Open circles:
 before baking, closed circles: after baking. Test temperature 6 K. }
 \label{BCPpower}
\end{figure}

\noindent For the EP samples (baked and unbaked) the exponents are
 found to be equal:
 $s=t=1.3 \pm 0.1$, see Figs.~\ref{EPpower1} and \ref{EPpower2}. This is consistent with a
 2-D model of a percolation-driven transition to coherent
 superconductivity
 ~\cite{lars2, Lars, sara}.
 For the BCP samples, however,
 the exponents $s=1.05 \pm 0.10$ and $t=1.4 \pm 0.1$ are different
 (see Fig.~\ref{BCPpower}), indicating a
 dimensionality of the percolating network which is slightly higher
 than two but still much smaller than three~\cite{straley}. The power
 law analysis points to different surface topologies and electronic
 structures of BCP and EP
 samples. As expected, baking has no influence on the topology, and hence does not modify the exponents.

\begin{figure}[phtb]
\centering
\includegraphics*[width=65mm]{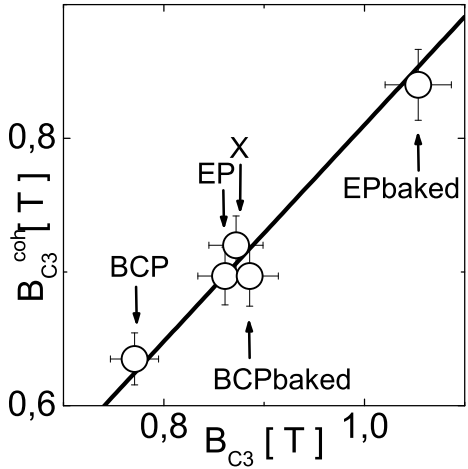}
\caption{Linear relation between $B^{coh}_{c3}$ and $B_{c3}$
obtained for samples with different surface treatment. The data point
labelled ``X'' refers to a Nb single crystal with an etched surface~\cite{Lars}.} 
\label{Hcc3}
\end{figure}

\noindent An interesting observation is that the coherent critical surface
field is always a fixed fraction of the Ginzburg-Landau-type
surface field: $B_{c3}^{coh}=(0.81 \pm 0.02)B_{c3}$, independent
of the surface topology, see Fig.~\ref{Hcc3}. This suggests some
intrinsic but yet unknown effect behind the formation of the
coherent surface superconductivity.

\subsection{Frequency dependence of the conductivity}
\begin{figure}[htb]
\centering
\includegraphics*[width=65mm]{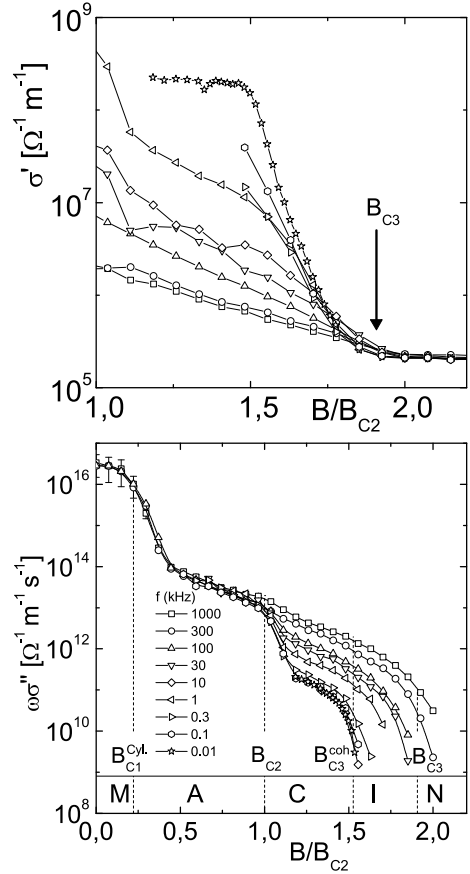}
\caption{The complex conductivity $\sigma(\omega)$ as a function of field for
 frequencies between 10 Hz and 1 MHz. The top graph shows $\sigma'$, the bottom graph $\omega \sigma''$. The indicated critical fields
$B_{c1}$, $B_{c2}$, $B_{c3}^{coh}$ and $B_{c3}$ separate five
phases: Meissner~M, Abrikosov~A, coherent surface superconductivity~C,
incoherent surface superconductivity~I, normal state~N.} \label{Cfreq}
\end{figure}

Up to now we have discussed the ac susceptibility and conductivity in
the quasi-static limit at $f=10~$Hz. The investigations have been
extended up to 1~MHz. In Fig.~\ref{Cfreq}, the quantity $\omega
\sigma''$, which
is a measure of the superfluid density, and the loss component
$\sigma '$ are shown as functions of the applied field for frequencies
between 10~Hz and 1~MHz. In the normal conducting regime above
$B_{c3}$ we find that $\sigma '=\sigma_n=(2.1\pm 0.3)
10^9~ (\Omega\mbox{m})^{-1}$ is frequency independent and also
field independent since it agrees with the value obtained in zero
field (see Sect.~\ref{bulk}). Below $B_{c3}$, $\sigma'$
becomes frequency dependent.

\begin{figure}[tbh]
\centering
\includegraphics*[width=100mm]{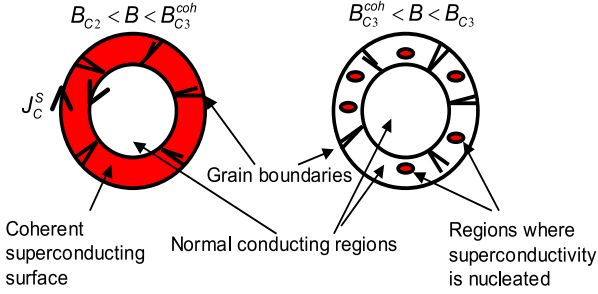}
\caption{Schematic view of the coherent and incoherent surface phases
 of a superconducting cylinder.}
 \label{conc}
\end{figure}

\noindent From the field dependence of
$\omega \sigma''$ it is possible to
distinguish five phases of the samples obtained by sweeping the dc
magnetic field~\cite{sara}. The indicated critical fields separate
the Meissner phase~(M), the Abrikosov vortex lattice phase~(A),
the coherent surface phase~(C), the incoherent surface phase~(I), and
finally the normal phase~(N). The two surface phases are depicted
schematically in Fig.~\ref{conc}.
The Meissner and Abrikosov phase are frequency independent.
The strongest frequency variation is observed close to the coherent
surface field
$B^{coh}_{c3}$.

\subsection{Surface critical currents}

 As a further evidence for the coherence of the surface
 superconductivity below $B_{c3}^{coh}$, a finite critical current
 has been detected~\cite{lars2, Lars, sara}. To this end, a
 small longitudinal gradient in the dc magnetic field $B$
 of the SQUID magnetometer has been utilized, the field inhomogeneity being $0.0025\%/$mm.
 During the motion of the sample through the pickup coil with a
 velocity $v_z$ the field gradient induces an azimuthal electric field
 along the circumference of the cylinder of radius $a$
$$E_{\phi}=\frac{a v_z}{2}\cdot \frac {dB}{dz} \;.$$

\begin{figure}[tbh]
\centering
\includegraphics*[width=65mm]{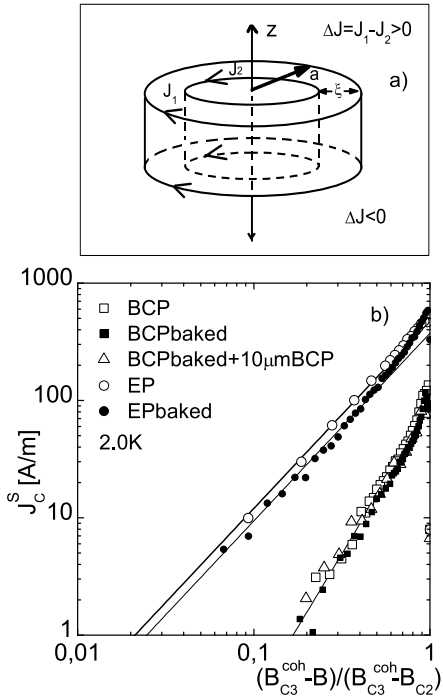}
\vspace{-5mm}
\caption{a) Schematics of the surface supercurrents according to
Fink and Barnes ~\cite{finkbarnes}. b) Dependence of the surface
current per unit length $J^{s}_c$ on the normalized magnetic field
$(B^{coh}_{c3}-B)/(B^{coh}_{c3}-B_{c2})$ for all samples.
($T=2~$K).} \label{Jsc}
\end{figure}

\begin{figure}[tbh]
\centering
\includegraphics*[width=70mm]{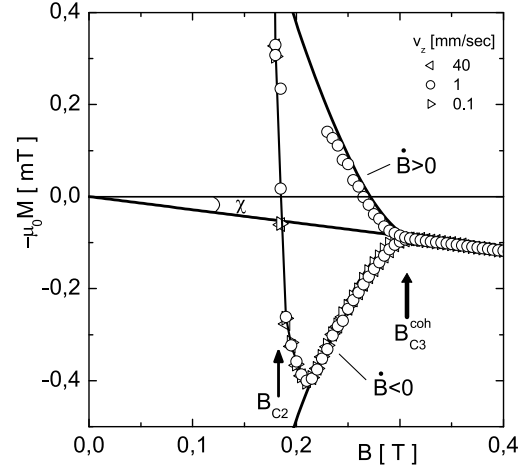}
\caption{Magnetization of an EP cylinder between $B_{c2}$ and
$B^{coh}_{c3}$ at 6.0~K measured along positive and negative field
gradients. Above $B_{c2}$, the magnetization is
independent of the scan velocity $v_z$. Note the small values of the
magnetization.} \label{Mc}
\end{figure}

\noindent Depending on the direction of motion, the response is either
 diamagnetic or paramagnetic.
 These responses are symmetric with
respect to the linear background $M_{el}(H)=\chi_{el} \,B/\mu_0$
due to the
 paramagnetism of the normal conduction
electrons in niobium (see sect.~\ref{paragm}). The
 induced surface current per unit length
\begin{equation}
J_{c}^{s}(B)=M-\chi_{el} B/\mu_0
\label{surfcurrent}
\end{equation}
is shown in Fig.~\ref{Jsc} for all samples as a function of the
reduced field $(B_{c3}^{coh}-B)/(B_{c3}^{coh}-B_{c2})$ at 2.0~K.
An important observation is that $J_{c}^{s}(B)$ does not depend on
the scan velocity $v_z$, which has been varied between 0.1~mm/s and
40~mm/s (see Fig.~\ref{Mc}). This indicates that $E_\phi$ is
always large enough to excite the critical surface supercurrent
density in the cylinder, $J_c^{s}(B)$.

\noindent According to Abrikosov~\cite{abrikosov} the surface current density
(current per unit length) should obey a power law
\begin{equation}
J_c^{s}(B)\propto \left(1-\frac{B}{B_{c3}}\right)^{v}
\end{equation}
 with $v=1.5$. For the EP samples we find an exponent $v=1.6 \pm 0.1$, consistent
with the Abrikosov calculation. The measured currents per unit
length of $477\pm 24$~A/m (unbaked) and $374\pm 24$~A/m (baked)
are in good agreement with predictions by Fink and
Barnes~\cite{finkbarnes} who consider a multiply connected
surface sheath with two currents flowing in the opposite direction
as illustrated by Fig.~\ref{Jsc}a. In this sense, our surface current~(\ref{surfcurrent}) must be interpreted as the difference between two large counterrotating currents. We observe a clear difference
between electropolished and chemically etched samples: the BCP
cylinders have a factor of six smaller critical surface currents
and the exponent is larger, $v=2.5\pm 0.3$.

\noindent Following Ref.~\cite{lars2} we compare the
exponent $v$ with the scaling prediction for a multidimensional
percolation network, $v=t-\Delta$~\cite{deutscher}. For the EP
sample the magnitude of the so called {\it twistedness index} of the
macrobond, $\Delta=t-v=-0.3\pm 0.3$ is in agreement
with the small and negative value found for 2-D granular
superconducting PbGe films~\cite{deutscher}. The higher value of
$\Delta$ for the BCP samples indicates that the surface currents have
to follow more complicated orbits than in EP samples.
The lower surface current densities $J^{s}_c$ in BCP-treated
surfaces
and the higher twistedness index $\Delta$ are possibly
related to the larger roughness at the grain boundaries.

\noindent The surface currents per unit length, as measured in our
experiment, are three orders of magnitude lower than the currents
that would be needed to excite a niobium cavity to rf magnetic
fields beyond $H_{c2} \approx 3\cdot 10^5~$A/m. 
However, our quasi-static measurements ($f=10$~Hz) do not exclude the possibility that much higher surface currents might exist in the non-equilibium state of superconductivity which is relevant for high-frequency cavities.


\section{Paramagnetic susceptibility} \label{paragm}

\begin{figure}[htb]
\centering
\includegraphics*[width=68mm]{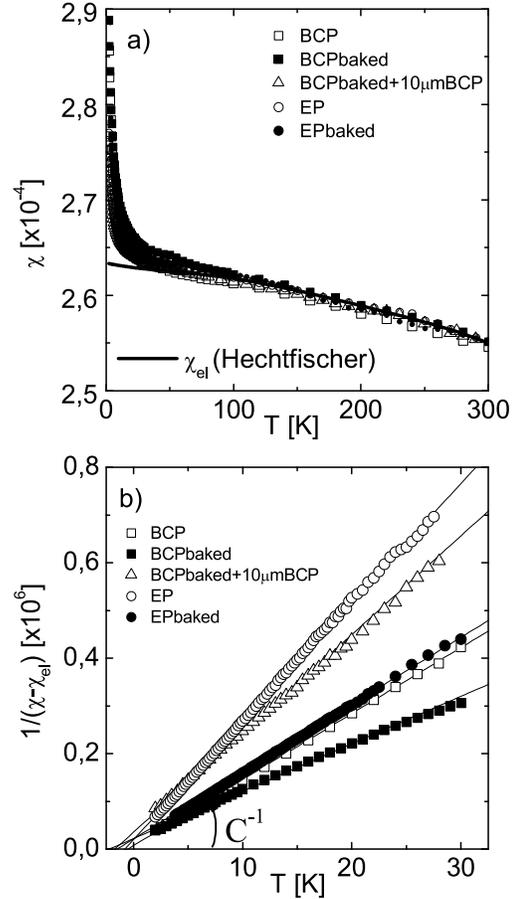}
\caption{a) Temperature dependence of the susceptibility of all
 samples measured at $B=700~$mT. b) Curie-Weiss contribution to the
paramagnetic susceptibility. Solid lines are fits to the
Curie-Weiss law $(\chi-\chi_{el})^{-1}=(T-\theta)/C$.}
\label{Curie}
\end{figure}

By measuring the magnetic susceptibility in a
large external field $B=0.7~$T where the entire sample is in
the normal state it is possible to search for magnetic impurities in
the niobium. The data are shown in Fig.~\ref{Curie}a for temperatures
between 2 and 300 K. Above $50~\mbox{K}$
our results on the paramagnetic susceptibility
$\chi_{el}$ of the normal conduction electrons in Nb
are in excellent agreement with previous measurements
 by Hecht\-fischer~\cite{hechtfischer}.
Below $50~\mbox{K}$ an additional contribution to $\chi$
is observed, which obeys a Curie-Weiss law
$$\chi-\chi_{el}=C/(T-\theta),$$
see Fig.~\ref{Curie}b. The fit values for the Curie constant $C$
and the Curie-Weiss temperature $\theta$ are listed in
Tab.~\ref{tcurie}. Such behaviour indicates the presence of
localized magnetic impurities. While $\theta\simeq -1~\mbox{K}$
for all samples, the Curie constant, being proportional to the
concentration of the localized magnetic moments,
increases during the bakeout.

\begin{table}[hbt]
\begin{center}
\caption{Curie constant $C$ and Curie-Weiss temperature~$\theta$.} 
\vskip 0.2truecm
\begin{tabular}{|l|c|c|}
\hline Sample & $C~[\mu\mbox{K}]$ &$\theta$~[K] \\
\hline BCP &$72.3\pm 0.1$ & $-0.5\pm 0.2$ \\
\hline BCP baked &$100.6\pm 0.07$ & $-2.2\pm 0.1$ \\
\hline EP & $40.2\pm 0.3$ & $-0.8\pm 0.2$ \\
\hline EP baked &$71.0\pm 0.1$ & $-1.5\pm 0.2$ \\
\hline BCP baked+$10\mu m$ BCP &$48.3\pm 0.4$ & $-1.7\pm 0.3$\\
\hline
\end{tabular}
\label{tcurie}
\end{center}
\end{table}

\noindent Oxygen vacancies in the Nb$_2$O$_5$ sheath would be a good
candidate for localized magnetic moments ~\cite{cava}.
Their Curie-Weiss behaviour has been observed in
Nb$_2$O$_{5-\delta}$ crystallographic shear structures with
$C\simeq 10~\mbox{mK}$ for $\delta\approx 0.17$~\cite{cava}.
Assuming the shear structure to be present in the
$d \simeq 5~\mbox{nm}$ thin Nb$_2$O$_5$ layer, one expects a
Curie constant $C=3 d/a \cdot 10~\mbox{mK} \approx
0.12~\mu\mbox{K}$. Since this is more than two orders of magnitude
smaller than the values in Tab.~\ref{tcurie}, oxygen vacancies
in the Nb$_2$O$_5$ sheath cannot account for the observed magnetic behaviour.

\noindent The paramagnetism of other impurities, which may be introduced
by the BCP and EP processes, like N, C, F, P, S, the hydrogen
bonded H$_2$O/C$_x$H$_y$ (OH)$_z$~\cite{halbritter} and some
niobium suboxides NbO$_x$ ($x \lesssim 1$), has not yet been
investigated. Perhaps they form clusters with large paramagnetic
moments. After baking, the Curie constant $C$ is found to be
increased by about $40-50~\%$. One possibility would be that
during baking additional magnetic moments are released from
external or internal surfaces. \\

\noindent The observation that a BCP-treated sample has a significantly larger
 Curie constant than an EP sample indicates that the magnetic moments
are not confined to the surface only but reside
 also deeper in the
material, probably in the grain boundaries which are more
pronounced in BCP than in EP surfaces.

\newpage
\section{Summary and conclusions}
As expected, the bulk properties of the niobium samples, $T_c$,
$B_c$, $RRR$ and $B_{c2}$, remain invariant when different
surface treatments such as chemical etching and electropolishing
or a low-temperature bakeout are applied. In contrast to this, the
superconducting properties of the surface itself are found to be
strongly modified by these treatments. Evidence for surface
superconductivity at fields exceeding the upper critical field
$B_{c2}$ of the bulk is found in all samples. The critical
surface field $B_{c3}$ is always larger than the value $B_{c3}=1.695
B_{c2}$ derived from the Ginzburg-Landau theory, the ratio
$r_{32}=B_{c3}/B_{c3}$ amounts to 1.86 for BCP samples and 2.1 for
EP samples. It increases further by baking the sample at $120 -
140^\circ$C for 24 to 96 hours. We interpret this enhanced surface
field as being due to increased impurity contents of the
niobium in a layer close to the surface and, related to this, a
reduced electron mean free path. The most likely contaminant is oxygen.

\noindent A most remarkable
observation is that two different phases of surface superconductivity
exist which are separated by a ``coherent'' critical surface field
 $B_{c3}^{coh}$: a coherent phase~C for applied fields between $B_{c2}$ and
$B_{c3}^{coh}$ with bipolar shielding currents going around the
whole cylindrical sample, and an incoherent phase~I between
$B_{c3}^{coh}$ and $B_{c3}$ which is characterized by disconnected
superconducting regions with normal zones in between. Both
$B_{c3}^{coh}$ and $B_{c3}$ depend on the surface preparation but
the ratio $B_{c3}^{coh}/B_{c3}$ has the value 0.81 for all
samples: BCP, EP, unbaked and baked. A power-law analysis of the
complex conductivity and resistivity reveals that at
$B_{c3}^{coh}$ a phase transition takes place between coherent and
incoherent surface superconductivity. For the EP samples the
exponents are in agreement with the expectation for percolation
through a two-dimensional network of superconducting and
resistive sections.
A different behaviour is seen in the BCP samples, here the dimensionality of the network
would have to be slightly larger that two. We suspect that this may be
related to weak links at the grain boundaries and
to more complicated, nonplanar current paths in the surface layer.

\noindent In the coherent phase C, a small net current around the Nb cylinder can
be induced by a time-varying magnetic field whose direction depends on
the sign of $\dot B$ according to Lenz's rule. However, this net
current, being the difference of two counter-rotating currents,
 is only a few 100 A/m and thereby
three orders lower than the rf currents
that would be needed to
operate an accelerating cavity at rf magnetic fields above
 $H_{c2} \approx 3 \cdot 10^5~$A/m.
Our steady-state results do not exclude the possibility that higher
surface currents might exist in non-equilibrium states, for
instance in the high-frequency fields applied to accelerator
cavities. In the BCP samples the surface currents are a factor of
six lower which again points to weak links at grain boundaries.

\noindent An important result of our investigations is that the various
 surface preparation
steps improving cavity performance have all a well-measurable
influence on the magnetic properties of the samples. Electrolytic
polishing of a BCP sample raises the critical surface field
$B_{c3}$ by about $ 10~\%$. The low-temperature baking leads to
 a further enhancement by
 $20\%$. The critical exponents of the power law fits near
$B_{c3}^{coh}$ are different for EP- and BCP-treated samples: we
get $s=t=1.3\pm 0.1$ for EP and $s=1.05\pm 0.1$, $t=1.4 \pm 0.1$
for BCP samples. From this one can conclude that the smooth EP
surface is able to support planar (two-dimensional) surface
currents while the rough grain boundaries in a BCP surface enforce
more complicated current patterns. The EP samples feature a
coherent surface phase which resembles the Meissner phase in the
bulk. In the BCP samples this coherent phase is disturbed by weak
links at the grain boundaries. 

\noindent The paramagnetic susceptibility in
the normal-conducting regime is dominated by the normal conduction
electrons in niobium but at low temperatures an additional
contribution is observed which obeys a Curie-Weiss law and can be
attributed to paramagnetic impurities. Their number
 is increased by low-temperature baking (LTB).
This in qualitative accordance with the generally accepted interpretation that LTB leads to
a partial reduction of the Nb$_2$O$_5$ layer and an oxygen diffusion
into deeper layers. Remarkably, the density of paramagnetic impurities
is larger in BCP than in EP samples, which may be an indication that
the impurities in grain boundaries play an important role. 

\noindent In our view, the above observations are of high
 relevance for rf cavities and underline the superiority of
 electropolished surfaces.
It must be emphasized, however, that
 measurements on niobium samples do not render cavity tests
 superfluous. One essential difference is that the samples are
 investigated in a large dc background magnetic field with a
 superimposed small ac field while in the cavities the rf magnetic
 field assumes large amplitudes. It is by no means obvious that the
 superconductor responds in the same way to these different
 conditions. In particular, the dramatic improvement in the
 high-field performance of EP cavities by applying the
 low-temperature bakeout could certainly not have been predicted from
 the $20\%$ growth of $B_{c3}$ observed in EP-treated Nb samples.
 According to our understanding the underlying mechanisms of the bakeout effect are not yet
 fully understood.
\section*{Acknowledgements} 
Thanks are due to R. Anton (University of Hamburg) for his advice and the possibility to use his equipment for sample preparation, to N. Steinhau-K\"uhl (DESY) for carrying out the BCP and EP treatments of the samples, to D. G\"orlitz (University of Hamburg) for his help and to W. Gil (University of Hamburg) for SEM and AFM measurements on the samples. 
The work of S.C. was supported by the Deutsche Forschungsgemeinschaft through Grant No. CA 284/1-1.

\section*{Appendix}
The SQUID-Magnetometer ${\rm MPMS_2}$ made by Quantum Design is
an excellent apparatus to measure the dc magnetisation and the
ac susceptibility of samples with a very high sensitivity and over a
wide range of magnetic fields and temperatures. A superconducting
solenoid produces a dc field up to $\pm
1\,{\rm T}$, and a copper coil with 8 turns produces ac-fields from $10\,{\rm nT}$ up to
$0.5\,{\rm mT}$, with frequencies from $0.01\,{\rm Hz}$ up to $1\,{\rm
kHz}$. The temperature can be varied from $2\,{\rm K}$ up to
$350\,{\rm K}$. The measurable range of
magnetic moments extends up to $\pm 0.3~$Am$^2$ with a sensitivity of
$10^{-10}~$Am$^2$.

\begin{figure}[htb]
\centering
\includegraphics*[width=80mm]{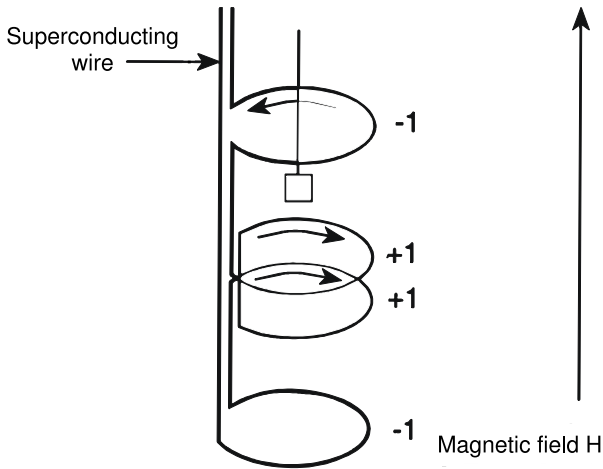}
\caption{Schematic coil configuration of the SQUID magnetometer.}
\label{gradiometer}
\end{figure}

\noindent The detection coil system is built as a gradiometer of
second order and made from a single length of superconducting wire.
The center coil with two turns is wound
clockwise, the upper and lower coils are $14\,{\rm mm}$ away
from the center coil and consists of a single turn each wound counter-clockwise (see Fig.~\ref{gradiometer}). This
configuration cancels noise due to fluctuations in the large magnetic field of the
superconducting magnet. Together with a transformer which couples the current
changes to the rf-SQUID the pick-up-coil
forms a closed superconducting loop.

\noindent For dc magnetisation measurements, the sample is moved through the
pickup coil at constant speed. The scan length is typically
$40\,{\rm mm}$, measurements are taken at 30-50 positions. At each
position the motion is shortly stopped for the measurement. The
current induced in the pickup coil
 by the moving sample is proportional to the
magnetic moment of the sample.

\noindent The ac susceptibility is measured at two
different positions: in the
center coil of the pickup system and in the lower compensation
coil. Before applying the ac field
 a nulling procedure is carried out to remove the induced signal from
 switching on the superconducting coil.\\

\begin{figure}[htb]
\centering
\includegraphics*[width=80mm]{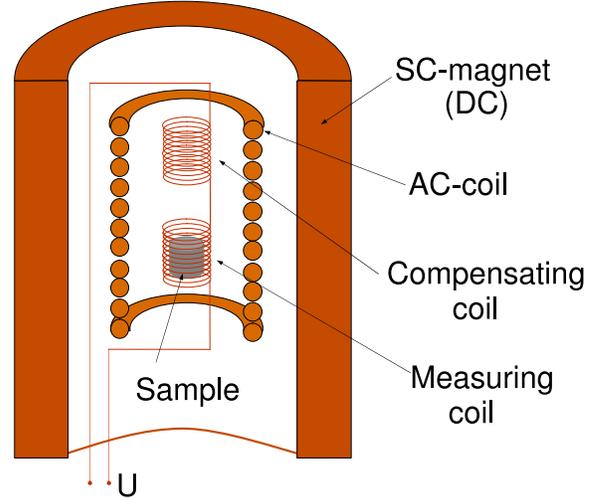}
\caption{Coil configuration of the mutual inductance magnetometer.}
\label{spulenengl}
\end{figure}

\noindent In the second magnetometer, sketched in Fig.~\ref{spulenengl}, the
measuring coil system consists of two identical coils
wound in opposite direction. Without sample the induced voltage is
almost balanced to zero. The sample is positioned in one of the two
pickup coils so that the measured voltage is given by the following
relation:
\begin{equation}
U= i I_{ac} \omega M q_f \chi
\end{equation}
where $I_{ac}=I_0 \exp(i\omega t)$ is the current in the ac coil,
$M$ the mutual inductance coefficient between the ac coil and one
of the pickup coils, $q_f$ the filling factor given by the ratio
of sample volume to coil volume and $\chi$ the ac susceptibility
of the sample. Since the two pickup coils are not exactly
identical there will be a background voltage $U_b$, that depends
on the frequency and amplitude of the applied ac field.
\begin{eqnarray}
U'&=& \cos\phi\,(U_m'-U_b')+\sin\phi\,(U_m''-U_b'')\nonumber \\
U''&=& -\sin\phi\,(U_m'-U_b')+\cos\phi\,(U_m''-U_b'')\nonumber,
\end{eqnarray}
where $U_m'$ and $U''_m$ are the real and imaginary part of the
measured signal. The system is calibrated with a spherical
GdCl$_3 \cdot6$H$_2$O sample of 3 mm diameter. For each ac
amplitude and frequency
the phase $\phi$ and the mutual coupling $M$ are determined from the
conditions $\chi''=0$ and
 $\chi'=C/T$, where $C=0.65\pm 0.01~\mbox{K}$ is the known Curie
 constant of GdCl$_3 \cdot6$H$_2$O .



\end{document}